\documentclass{nime-alternate} 

\usepackage{anonymize} 

\usepackage{csquotes}
    
    \DeclareQuoteStyle[american]{english}
        {\itshape\textquotedblleft}
        [\textquotedblleft]
        {\textquotedblright}
        [0.05em]
        {\textquoteleft}
        {\textquoteright}
    
\begin{document}
%
\conferenceinfo{NIME'19,}{June 3-6, 2019, Federal University of Rio Grande do Sul, ~~~~~~  Porto Alegre,  Brazil.}
\title{Ephemeral instruments}

%
%
%
%
\label{key}
%

\numberofauthors{1} 
%
\author{
%
%
\alignauthor
\anonymize{Vincent Goudard}\\
       \affaddr{\anonymize{Sorbonne Université}}\\
       \affaddr{\anonymize{LAM, Institut Jean Le Rond d'Alembert}}\\
       \affaddr{\anonymize{Institut de Recherche en Musicologie}}\\
       \affaddr{\anonymize{Paris, France}}\\
       \email{\anonymize{vincent.goudard@sorbonne-universite.fr}}
}



\maketitle
\begin{abstract}
This article questions the notion of ephemerality of digital musical instruments (DMI). Longevity is generally regarded as a valuable quality that good design criteria should help to achieve. However, the nature of the tools, of the performance conditions and of the music itself may lead to think of ephemerality as an intrinsic modality of the existence of DMIs. In particular, the conditions of contemporary musical production suggest that contextual adaptations of instrumental devices beyond the monolithic unity of classical instruments should be considered. The first two parts of this article analyse various reasons to reassess the issue of longevity and ephemerality. The last two sections attempt to propose an articulation of these two aspects to inform both the design of the DMI and their learning.
\end{abstract}






\section{A critique of longevity}
\subsection{DMI will survive}
The issue of DMIs longevity is a complex one that has been raised several times in the NIME\footnote{New Interfaces for Musical Expression.} literature (and related) and has been a growing topic of discussion in the last decade \cite{baguyos_contemporary_2014}\cite{morreale_design_2017}. The authors of these papers identified a number of causes of this situation, ranging from technical, methodological, musical or sociological, and came up with insights and directions to address this issue, like new frameworks for building and evaluating instruments  \cite{jorda_digital_2004} \cite{morreale_design_2017}, better documentation \cite{bonardi_preservation_2008}, new pedagogical methods and forms of community-building \cite{marquez-borbon_problem_2018} as well as trying to establish musical notation and repertoire for these new instruments \cite{mamedes_composing_2014}\cite{mays_notation_2014}. However, in the majority of these papers, the lack of longevity of DMIs is essentially considered as a defect, or at least a problem to solve.

As early as 1975, electroacoustic music composers at GRM were thinking about the preservation questions raised by a music \enquote{written on sand} \cite{chion_musique_1977}. Some composers said that they did not care about it and made their music for the present time, while others saw in the emerging digital era the possibility of preserving their work for the future. As we know today, trading sand for silicon (or cloud for that matter) has not totally solved the problem.

Since DMIs have largely integrated the very notion of musical work, sometimes intrinsic to instruments, the desire to preserve musical pieces has been partly transposed into the concern to preserve the instruments and tools used in their production. But what are the reasons for this pursuit of longevity? And what legitimates the fact that the longevity of an instrument is acknowledged as a quality criterion? 

The desire for longevity is ontologically linked to a reaction deeply rooted in our condition of mere mortals, that consists in seeking a way to ensure our survival, especially by the transmission of knowledge and the creation of traditions. Leroi-Gourhan analysed the phenomenon of traditions as a way to externalize and transmit our memory as “action sequences” \cite{leroi-gourhan_geste_1964} and Stiegler further built on this idea by calling “grammatization” the process whereby the temporal continuum of human behaviours is turned to a spatial discreteness, that allows to embed them in tools \cite{stiegler_for_2010}.

Humans have thus developed methods and tools such as the psalmody of texts (especially religious), or writing as a means both of recording information for later use and of transmitting knowledge to those who survive them. Writing has partly freed humans from the need for oral tradition by transferring this knowledge to a physical medium, which in turn allowed to capitalise and speculate on knowledge.

Hence, the notion of longevity crosses the field of the arts and sciences, at the borders of which musical instruments stand. In the history of art, mainly the durable works remain, engraved in the stone from which sculptures are made. Similarly, science aspires to find sustainable laws to describe the observable world, that are most of the time based on the perennial language of mathematics. But, if obvious longevity of a work often acts as an asset for its own legitimation, when it comes to digital instrument, even more so when it is conceived as an interactive mean to create a musical \textit{experience}, the matter cannot be resolved under the same conditions.
	
\subsection{Longevity, adoption, success}
Two aspects often seem to be confused: the longevity of an instrument on the one hand, and its “success” on the other. Moreover, the notion of success, symptomatic of our digital society that seeks to quantitatively evaluate all fields of reality, is eminently subject to the perspective adopted, but often seems to be considered as the adoption rate by a community of instrumentalists, beyond the financial aspects of commercial success.

These three \textit{aspects}, \textit{longevity}, \textit{success} and \textit{adoption}, are nevertheless quite different, partly independent or even contradictory. There are good examples of this discrepancy: Michel Waisvisz's \textit{The Hands} \cite{torre_hands:_2016} or Serge De Laubier's \textit{Meta-Instrument} \cite{couprie_meta-instrument:_2018} are two instruments that have had a durable existence (25 and 30 years, respectively), supported by regular practice by their inventors, without having yet been adopted by a large community of users. Conversely, the ephemerality of a tool does not systematically lead to a lack of popularity\footnote{Let us here consider all the ephemeral gadgets, which under the influence of a fashion and/or a powerful advertising campaign invade the market, or all the devices that become obsolete when a new device replaces them, such as smartphones which, in addition to replacing our old phones, have swept away audio players, GPS, portable game consoles, flashlights, etc. in one go.}, and even less to a lack of musical interest in the performances made with it.

The notion of success therefore depends on the point of view adopted, whether it be that of luthiers who create instruments for others, or that of ones who create instruments for their own selves. In the latter case, the customisation of the instrument to one's own needs and aesthetics may prove to be such that it is difficult for others to adopt it.

Also, technical evolutions as well as fashion trends can lead to the reappearance of an instrument once fallen into oblivion. Here, we can appreciate the insight of François-Alexandre Garsault, quoted by Malou Haine in \cite{haine_les_2018}, who in his \textit{Division of instruments according to their different uses} (1761) classified a series of instruments, including the harp and the \textit{guitarre} (sic), in the category of \enquote{Instruments out of use, but that can come back}.

\subsection{Longevity versus stability}
The question of the sustainability of an instrument implicitly raises the question of its historical stability. As such, the organological history of European musical instruments reveals many factors that lead to the appearance, evolution or disappearance of musical instruments. In particular, the many technological innovations during the industrial revolution are interesting because this well-documented period illustrates the beginnings of major revolutions that would occur in the 20th century, while raising the very question of the form stability of instruments. So, when the traverso was equipped with the keying system invented by Boehm in 1832 and became a Western concert flute, was it a new instrument? At what point do we decide that an instrument that is undergoing changes is no longer the same?

\section{Ephemerality everywhere}
\subsection{In the sound}
First of all, let us recall an obvious fact: music itself is inherently intangible, evanescent and requires sustained energy to last. The profound nature of the sound phenomenon stays in \emph{permanent ephemerality}. Music, in its sensible form, only exists during the time of its performance. While the instruments used to produce it can be durable, their convocation and the sound itself are always temporary. 

\subsection{In the performance context}
\noindent Even when notated on a score, music as a performing art is in constant reinterpretation. This interpretation makes it possible to transform a score notated in symbolic form into a sensible expression subject to variations. It may be objected that this interpretation only exists when the music is notated in a symbolic way, leaving room for the interpreters to play it their own way in the performance context. But is it still the case when music is “fully notated” down to the sound itself, as is the case on an audio record? Does this mean that the interpretation disappears? The performances of live spatialization of electroacoustic music by trained musicians or the various practices of remix found in hip-hop tend to prove the opposite. Any musical performance, even the mere playing of a record, inevitably calls for a new listening context, as it necessarily occurs in a unique present moment. Between the recorded sound and its listening, we find the same \emph{differance}\footnote{\textit{Differance} is a concept proposed by Derrida \cite{derrida_lecriture_2014} to refer to both the adjournment and the differentiation taking place between a text and its meaning.} as between a score and its performances.

\subsection{In the score}
\noindent  Musical scores are partly integrated into digital musical instruments, for which Schnell and Battier coined the term \emph{composed instruments} in \cite{schnell_introducing_2002}. The score was itself subject to a more open re-configuration since the middle of the 20th century and gradually integrated algorithmic possibilities into its creative process: dynamic and interactive processes put the stability of note figures into motion. Several composers\footnote{ Among those who composed and analysed dynamic scores, see the works of Hajdu \cite{hajdu_disposable_2016}, Bhagwati \cite{bhagwati_vexations_2017} or Freeman \cite{freeman_extreme_2008}.} thus question the stability of the score by using the computer to create ad-hoc instances, either through generative algorithms or by introducing improvised parts into hybrid forms for which Dudas proposed the term \emph{comprovisation} \cite{dudas_comprovisation:_2010}. Is it so, that digital technology offers that ideal medium that would allow both the preservation of musical works as much as their mutation?

\subsection{In technology}
The materials used for acoustic instruments seem to age relatively well. Electronic hardware ages poorly in comparison, and the copper of its circuits is more fragile than that of brass instruments. Moreover, the extreme miniaturization of microprocessors often makes them impossible to repair; they need to be replaced and there is great chance that the substitutes will be new, different versions. Computer code, in its compiled form, is just as cryptic as the microprocessor: an unreadable block that embodies the paradox of computer-based notation as compared to traditional paper—we are writing things which we can no longer read. And when the operating system is updated, chances are it will no longer be able to read them either.\\
In an article where he compares the ontological differences between hardware and software, Nicolas Collins \cite{collins_semiconducting_2013} summarizes their relation to time with the formula: \enquote{hardware is yesterday, software is now}, what could be translated as the fact that software is under permanent update while hardware is ever outdated. Neither seems to be able to offer a reliable continuity between the past and the future.
	
\subsection{In the economy}
In addition to the obsolescence of technology, DMIs are confronted with the effects of consumer society. For more than a century, the industry has increasingly promoted a disposable paradigm by encouraging consumers \enquote{to trade for style, not just for technological improvements} \cite{slade_made_2006} and while organizing planned obsolescence.

This economic model also affected that of the performing arts, which promotes creations much more than the revival of a show to such an extent that, as Georg Hajdu recalls in \cite{hajdu_disposable_2016}: \enquote{Pieces rarely see more than a single performance}. Artist residencies are likewise targeted towards new creations and rarely propose the continuation of previous works.

This economy of obsolescence (planned or not) does not favour attachment to an instrument and, as far as commercial MIDI controllers are concerned, the cheap plastic they are most often made of degrades the value that can be attributed to a traditional acoustic instrument. The attachment and commitment to a virtual instrument is also challenged by its virtual nature. Most commercial software is now moving towards a rental—rather than purchase—economy, since the purchase no longer guarantees the sustainability of the property.
	
\subsection{In the instrument}
The musical instrument is also, as Bernard Sève points out in \cite{seve_instrument_2013}: \enquote{an unstable compromise between non-convergent qualities}. For acoustic instruments, this compromise between gestural ergonomics and acoustic performance, imposed by the physicality of the materials, is generally fixed in a fitted and glued assembly. This bonding acts as a stabilizing factor compared to a digital environment in which the absence of physical constraint leaves the instrument open-hearted, ready to be modified at any time.

Bill Buxton pointed out the difference between standard, military and artistic specifications to underline the higher requirement of the latter \cite{buxton_artists_1997}. Art-driven design require great finesse indeed. Tuning the sonic and ergodynamic\footnote{Magnusson proposed this term in \cite{magnusson_ergodynamics_2019} to name the \enquote{expressive power and depth of an instrument}.} qualities of a musical instrument is a quest for an inframince\footnote{Duchamp \cite{duchamp_notes_2008} coined the term \textit{inframince} in a series of examples depicting a difference so small that it can only be imagined.} for which there is no agreed specifications. But another particularity of the technologies used for live performance is that they are \enquote{devoted to an experience, not a sound track; unavailable for reshuffle or back-up or exchange or duplication}, as Nicolas Collins notes in \cite{collins_why_2008}.

Thus, the sustainability of the instrument outside the very duration of performance is not an essential criterion and it is not uncommon for digital musicians\footnote{Andrew Hugill defines a digital musician in \cite{hugill_digital_2019}, underlying the fact that they \enquote{not defined by their use of technology alone”, but also have “a certain curiosity, a questioning and critical engagement that goes with the territory}.} to modify their instrument minutes before the beginning of a concert, just for the needs of the present moment.
	
\subsection{In the accidents}
Indeed, the risk of dysfunction is not a major obstacle to many musical performances. Bugs and artefacts caused by malfunctions are proving to be fertile sources of musical materials and subverting the cryptic functioning of processors reveals an invisible aspect of them, bringing to the surface their very nature, beyond the purpose for which they were designed\footnote{Among significant examples, Yasuano Tone's works on “wounded CDs”, Nicolas Collins's works on dead circuitry or Carsten Nicolai's sonification of raw data exemplifies this approach.}. David Zicarelli, quoted by Cascone in \cite{cascone_aesthetics_2000}, sums it up in these terms: \enquote{I would only observe that in most high-profile gigs, failure tends to be far more interesting to the audience than success}.

\subsection{No more need for tradition ?}
The appearance of musical notation made performance no longer necessary for the only purpose of transmission; audio recording made performance no longer necessary for the only purpose of listening; computers and sound banks made the learning of a particular instrument no longer necessary to produce the sound of that instrument\footnote{See for example, Stravinsky's Rite of Spring edited by Jay Bacal with VSL: \url{https://youtu.be/PB3njyDW8SY}.}; and now, artificial intelligence makes the very act of composing no longer necessary for music to be composed\footnote{See for example the outcomes of the FlowMachines project by François Pachet et al. in \cite{hadjeres_deepbach:_2016}: “Daddy's car” and “DeepBach” (\url{https://youtu.be/LSHZ_b05W7o}, \url{https://youtu.be/QiBM7-5hA6o}).}.

In 1964, Leroi-Gourhan, who saw in the computing machine the unprecedented possibility of outsourcing memory, was wondering what would happen if the machines became capable of \enquote{writing perfect plays, creating inimitable paintings} \cite{leroi-gourhan_geste_1964}. In 1992, John Cage seemed to answer him a radical way, when saying: \enquote{We don't have to have traditions if we free ourselves from memory} \cite{sebestik_ecoute_1992}. 

However, if it is possible to evolve, as Buci-Glucksmann describes it in \cite{buci-glucksmann_esthetique_2003}, from a culture of the object to a culture of flows, she remarks that in a country like Japan that values impermanence positively, the ephemeral has a central place while being deeply rooted in tradition.	

The resolution of this apparent antagonism between Cage's position and Buci-Glucksman's seems to lie in the displacement of objects (or flows, for that matter) supported by tradition, in the reformulation of the motivations for sustainability and ephemerality.

\section{Coupling the perennial and the ephemeral}
\subsection{DMI as a weird and wild assemblages}
The very term DMI, which has gradually invaded the NIME literature, may lead us to believe that it is a well-defined category when it is in fact a hodgepodge of objects only sharing their use of digital computation. A resulting bias in the assessment of the failure of DMIs to reach maturity stems from the fact that a musical instrument is often still considered as a coherent whole, requiring longevity, similar to the acoustic instruments taken as role models.

Yet, the modularity induced by electronics and digital technology has atomised the integrity of the instrument. This atomisation can both be understood in the sense of “destructed” but also in the sense of “fragmented into atomic bits”. On stage, we can further observe that DMIs are often fragile, prototypical assemblies, full of cables ready to be interchanged minutes before the concert, or even during it. So why should we consider DMIs as durable monoliths rather than ephemeral assemblages\footnote{Following Deleuze and Guattari's concept proposed in \cite{deleuze_mille_1980}.} that they most often are?

From this point of view, the academic format of a conference such as NIME makes it difficult to present DMIs in their chaotic form and their selection is biased by the fact that their authors often belong to the academic world. This favours a presentation of duly considered technical criteria rather than the presentation of hectically connected algorithms whose functioning is not really understood, except for the fact that the musician who plays them does wonders.

By confronting an ephemeral instrumental setup, the instrumentalist, however virtuosic they may be, necessarily finds themselves in tension with a wild instrument to tame. This calls for an intense gestural and auditory attention and the research of resonance with the instrument. (Otherwise, one might as well compose comfortably at home and provide the listener with an audio record to be played with a single button). Maybe more important than longevity, here is an interesting design criterion for digital lutherie: the possibility that the instrument spins out of control.

\subsection{Cooking instruments on the fly}
Another reason that contributes to the stability of acoustic instruments is related to their physicality and manufacture, which requires a considerable amount of work compared to the virtual arrangement of software blocks—it takes more than two months to build a cello for a luthier who knows his job! Conversely, Bowers and al. promoted the use of readymade objects as half-made infra-instruments \cite{bowers_not_2005} and “pin-and-play” ad-hoc instruments \cite{bowers_creating_2006}, while rethinking the life-cycle of an instrument with such kind of quickly-built ephemeral assemblages.

More generally, as one create a DMI with an audio programming environment, the software not only provides basic functions but comes with libraries, ready-made examples, supplemented by countless online resources, ready to be downloaded, copied and pasted.

This means that the building of a DMI can be a much faster subtractive process: rather than starting from a blank page, it is possible to search for a version close to what one wants to achieve and modify it from there. Nicolas Collins compared this simplicity to cooking, emphasising its democratisation: \enquote{What it means is that if you are doing live performance, if you do need specialized instruments, it's almost more like cooking than it is building musical instruments. Everybody cooks, you don't need to go to ‘chef schools’!} [Collins, personal communication].

Recent evolutions in audio-programming languages tend to address the issue of sustainability by creating Domain Specific Languages that can be exported to various targets. Hardware platforms like Bela\footnote{Bela: \url{https://bela.io}} or The Owl\footnote{The Owl: \url{https://www.rebeltech.org}} and languages like FAUST \cite{orlarey_faust_2008} or the announced SOUL language by Roli\footnote{Announced at the Audio Developer Conference 2018. \url{https://youtu.be/-GhleKNaPdk}} all reflect this trend. It is worth noting that FAUST, which was designed with preservation in mind\footnote{FAUST was a key component of ASTREE, a project focusing on the preservation of musical work with electronics.}, actually helps building ephemeral instances by offering both an online compiler and just-in-time compilation.
	
\subsection{A three-way relationship}
If we therefore stop considering the ephemerality of instruments as a problem, we can consider how longevity and ephemerality can be articulated in the agency of DMI practices. It can be conceived as a tripartite coupling between materials, musician and context. Each component of this triad may then has a different degree of stability.

\subsubsection{The grand repertoire}
The components of a DMI can be considered as belonging to a large repertoire of both material and immaterial heritage. The material repertoire includes any physical material that can be used in the construction of acoustic instruments: raw materials as well as manufactured, machined or mechanical parts. 

The immaterial repertoire is all the theoretical knowledge and cultural heritage one can rely on during the making of an instrument\footnote{Obviously, practical knowledge is also essential to the making of an instrument, although it does not really belong to the shareable heritage to which I refer here.}: music theory, scientific and technical knowledge, established playing techniques, musical repertoire, etc. This knowledge helps to shape the materials and imprint musical markers to create the instrument: the placement of frets, the tuning of strings, the layout of keys, etc.

In the case of DMIs however, the repertoire of physical materials is considerably expanded by reified knowledge available as digital materials, either in the form of computer code or datasets (e.g. audio samples, impulse responses, scores, etc.) that enables the musical qualities of an instrument to to be shaped beyond what is possible with physical materials.
This set, as heterogeneous as it may seem, constitutes a shareable repertoire from which digital luthiers can draw the necessary ingredients for the design of their instrument.

\subsubsection{The musician in-progress/in-process}
The second element of the assemblage is the musician\footnote{Here, the term “musician” stands for the blurring roles of instrumentalists, composers and luthiers.}. Musicians are alive and subject to change: their knowledge, feelings and desires, musical skills and awareness, projects and physical capacities, all evolve during their existence. This evolution is reflected in the instrumental device, by the addition or removal of features, or the development of new instruments related to a new musical project. Just like you may learn to ride with a bicycle equipped with side wheels and later remove them, DMI can offer evolutive assistance for progressive learning. A co-dynamic relation with one's instrument can help improve the intimacy between the musician and the technical object becoming an instrument.

\subsubsection{The hic et nunc context}
Eventually, the DMI can be adapted to the context of performance, which is generally more ephemeral than the two aspects mentioned above.

Expanding their own musical repertoire by drawing on the grand repertoire mentioned above and on their own experience, musicians selects a subset of elements in the perspective of a particular performance, for a singular artistic proposition, and to meet the spatial and temporal conditions of the performance, as well as the audience. As an example, the costless duplication of code offers possibilities to rescale DMIs from soloist to collective instruments by distributing control over multiple interfaces. New projects can imply starting from scratch, but existing ones often only involve contextual adjustments rather than a thorough reprogramming of one's system. Kiefer and Magnusson coined the term “pre-gramming” \cite{kiefer_live_2019} to describe that particular kind of preparation.

\section{Playing the DMI}
As we can see, creating a DMI can be a very fast process as it can be done by simply assembling already pre-built elements. But once the assemblage is done, how do we learn to play it?

\subsection{Parallel composition}
\indent Traditional acoustic instruments are supported by methods and repertoire that can rely in turn on the instrument's stability. But for a new—possibly unique, possibly ephemeral—DMI, such resources are scarcely available. Software comes at best with manuals, but manuals usually explain how to make the software run, not how to play music with it.

From there, the learning process can follow two seemingly opposite directions: finding the right moves to play desired sounds and finding the right sounds for chosen gestures. A consequence is that often, learning a new DMI starts right from its conception and is a co-dynamic process that accompanies its development up to the \textit{pre-gramming} of the instrument, with back and forth between moments of play and moments of adjustment.

\subsection{Entering the future backwards}
DMIs and their practices integrate the know-how inherited from electroacoustic music since the middle of the 20th century. The pedagogy of electroacoustic music essentially developed a musical theory of listening \cite{schaeffer_traite_1966} and metaphors for composing \cite{bayle_musique_1993} but conceived at a time when electroacoustic music could only be composed, before real-time audio allowed live practices. As a result, these theories were more oriented towards musical composition than performance as such.

In the absence of established musical notation for sound, experimental electronic music is largely oriented towards free improvisation. This implies a letting go enabling the instrument to express its potentialities and a practice of “aurality”\footnote{Described by Savouret as a music theory for the audible.} to “enter the future backwards” \cite{savouret_introduction_2010} and react to what is coming out of the instrument rather than completely controlling it.

\subsection{Spotting the resonances}
	The learning of an instrument (beyond the learning of the heritage idioms of this instrument) thus requires a search for resonance. We can experience this resonance at an acoustic level, but more generally as an empathetic resonance, which consists in immersing ourselves in the instrument to find the spaces that will (re)sound satisfactorily, to find the “sweet spots” where what we hear meets what we were seeking—sometimes unknowingly. Since mathematical linearity is rarely satisfactory at a perceptual level, this exploration involving the coordination between play and critical listening is essential to adjust the mapping functions that will define the instrument's behaviour.
	
\subsection{Naming butterflies}
The musical exploration of a DMI brings out unknown musical forms, like ephemeral butterflies. Learning a DMI therefore often involve an entomologist-like task of pinning these sonic creatures and giving them a name. This naming will allow to come back to them later on (by saving them in presets for example) as well as to discuss with other musicians about a performance which, in the absence of established musical idioms on which to rely—like scales or time signature— can be cruelly lacking references. Such a task was led in the development of “John, the semi-conductor”, an open score system described in \cite{goudard_john_2018}.

\subsection{Practicing on stability}
While a DMI can be an unstable assemblage, its individual components may provide more stable grips. For example, if the performance is based on a written score, the instrumentalist can learn the sequence of appropriate gestures necessary for its realisation\footnote{A interesting and critical example is the piece Aphasia by Mark Applebaum (\url{https://youtu.be/wWt1qh67EnA}), where the performance relies on gestures and a soundtrack which are totally notated, yet to be performed.}.

As far as the behaviour of the DMI is concerned, one can partly transfer one's knowledge of other DMIs to a new instance one is trying to learn. For example, the integration of an FM synthesis into a DMI can help a person familiar with this type of synthesis to navigate its timbre space (bells, siren, brassy, wiggly, etc.), independently of the control interface plugged onto it, relying on their own knowledge and representation of FM synthesis parameter space. The timbre space of various audio syntheses can also be remapped on a common and more stable perceptive space (e.g. pitch, loudness, brilliance, etc.) that abstracts control from their differing parameter spaces, such as presented in \cite{wessel_timbre_1979}, \cite{arfib_strategies_2002}, \cite{schwarz_sound_2012} or \cite{tubb_divergent_2014}.

Likewise, an expertise can be acquired on a gestural interface, which calls for specific gestures and moves\footnote{For instance, consider the “launchpad” scene, characterized by the publication of battle-videos of rhythmic virtuosity.}; this expertise relies on an embodied spatial memory that—to some extent— remains partly independent from the audio syntheses or effects controlled with the interface. The behavioural stability of the instrument can also be of virtual nature, for example when using dynamic intermediate models \cite{goudard_dynamic_2011}, which can act as a stable reference taking place between various changing syntheses and interfaces.

Overall, this transposed and “modular” knowledge can only provide broad outlines of what is necessary for the subtle practice of an instrument. The devil's is obviously to be found in the details.

\subsection{A vessel for memory}
DMIs are heterogeneous vessels loaded with memories of our performing, composing or instrument-making experiences. The sounds that we collect, the synthesis algorithms that we develop (or download), the parameters that we adjust, the kitchen recipes and mapping functions that we carefully craft, all contribute to the evolution of a personal repertoire where ephemeral instances crystallize. Magnusson proposed the term epistemic tool to describe a musical instrument as \enquote{a designed tool with such a high degree of symbolic relevance that it becomes a system of knowledge and thinking in its own terms} \cite{magnusson_epistemic_2009}.

Thus, DMIs tend to be evolving assemblages of these stored memories and often involve activities that are not generally associated with instrumental practice, such as file management, bookmarking online resources or organizing sound banks, in order to be able to convene these resources as quickly as possible during the performance.

It is remarkable that the possibilities of duplication and dissemination offered by digital media and the Internet have not led to the standardisation of instrumentariums; digital musical instruments are often very personal and singular.

\section{Conclusion}
This article has presented how the ephemerality of DMIs should not only be considered as a problem, but as an intrinsic modality of their ontology. Rather than opposing longevity, it actually informs the technical design of the environments conducive to their development and sustainability.

Ephemerality of the tools does not prevent great music to be produced, nor great musical performances to happen. On the contrary, it can both help to adapt musical assemblages to contexts that are in essence ephemeral and to challenge human's ability to respond to a fleeting, untamed musical setup. In the end, great musical works seem to find their way, sustained by the care and the work of those who recognize these works as master pieces. These works may stand the test of time being dispersed, distributed, transformed, recomposed, reinterpreted or even renamed, by all those who will attach importance to them. This loving care probably belongs to the part of our memory that we cannot outsource in a tool and that redefines tradition and preservation outside the technological frame.

If digital technologies reach maturity one day, we may be able to rely on stable and sustainable instruments. In this case, following Garsault's premonition, it should not be forgotten to classify all the ephemeral instruments that preceded them in the category of \enquote{instruments out of use, but that could come back}.

\section{Acknowledgments}
This work was carried out within a doctoral program supported by the Collegium Musicæ, Sorbonne University.
All URLs in this document have been checked on April 9, 2019.

%
\bibliographystyle{abbrv}

	\bibliography{nime-references}

%
%

\balancecolumns 

\end{document}